\documentclass[lettersize,journal,compsoc]{IEEEtran}
\usepackage{amsmath,amsfonts}
\usepackage{algorithmic}
\usepackage{algorithm}
\usepackage{array}
\usepackage{caption}
\usepackage[caption=false,font=normalsize,labelfont=sf,textfont=sf]{subfig}%caption=false,
\usepackage{textcomp}
\usepackage{stfloats}
\usepackage{url}
\usepackage{ragged2e}
\usepackage{verbatim}
\usepackage{graphicx}
\usepackage{multirow} 
\usepackage{booktabs}
\usepackage{cite}
\usepackage{color}
\usepackage{xcolor} 
\begin{document}

\title{AniGaussian: Animatable Gaussian Avatar with Pose-guided Deformation}

\author{Mengtian~Li, Shengxiang~Yao, Chen Kai, Zhifeng~Xie$^{*}$, Keyu~Chen$^{*}$, Yu-Gang~Jiang, \IEEEmembership{Fellow, IEEE}
        % <-this % stops a space

\thanks{\textbullet\ $^{*}$: Corresponding author}
\thanks{\textbullet\  Mengtian~Li is with Shanghai University, Fudan University. E-mail: mtli@$\left\{shu \setminus  fudan \right\}$.edu.cn}
\thanks{\textbullet\  Shengxiang~Yao, Chen Kai and Zhifeng~Xie are with Shanghai University.  E-mail:$\left\{ yaosx033 \setminus zhifeng \_ xie \right\}$ @shu.edu.cn,myckai@126.com}
\thanks{\textbullet\ Keyu~Chen is with Tavus Inc. E-mail: keyu@tavus.dev.}
\thanks{\textbullet\ Yu-Gang~Jiang is with the School
of Computer Science, Fudan University. E-mail: ygj@fudan.edu.cn}
}

\markboth{Journal of \LaTeX\ Class Files,~Vol.~14, No.~8, August~2021}%
{Shell \MakeLowercase{\textit{et al.}}: A Sample Article Using IEEEtran.cls for IEEE Journals}

\IEEEpubid{0000--0000/00\$00.00~\copyright~2021 IEEE}

\IEEEtitleabstractindextext{

% \setcounter{figure}{0}
% \captionsetup{type=figure}
% \noindent\includegraphics[width=0.92\textwidth]{pic/fig1.png}
% \captionof{figure}{{\bfseries Infinite motion}: We propose a novel method for generating infinite motions, based on the timestamps featured in our HumanML3D-Extend dataset. This approach not only enables the generation of extremely long motions but also facilitates precise control over actions within specific time intervals.}
% \label{fig:1}

\vspace{1em}
\begin{abstract}
\justifying
Recent advancements in Gaussian-based human body reconstruction have achieved notable success in creating animatable avatars. However, there are ongoing challenges to fully exploit the SMPL model's prior knowledge and enhance the visual fidelity of these models to achieve more refined avatar reconstructions. In this paper, we introduce AniGaussian which addresses the above issues with two insights. First, we propose an innovative pose guided deformation strategy that effectively constrains the dynamic Gaussian avatar with SMPL pose guidance, ensuring that the reconstructed model not only captures the detailed surface nuances but also maintains anatomical correctness across a wide range of motions. Second, we tackle the expressiveness limitations of Gaussian models in representing dynamic human bodies. We incorporate rigid-based priors from previous works to enhance the dynamic transform capabilities of the Gaussian model. Furthermore, we introduce a split-with-scale strategy that significantly improves geometry quality. The ablative study experiment demonstrates the effectiveness of our innovative model design. Through extensive comparisons with existing methods, AniGaussian demonstrates superior performance in both qualitative result and quantitative metrics. 
\end{abstract}

}
\maketitle

\section{Introduction}
\label{sec:Introduction}
Creating high-fidelity clothed human models holds significant applications in virtual reality, telepresence, and movie production. Implicit methods based on occupancy fields ~\cite{saito2019pifu, saito2020pifuhd}, signed distance fields (SDF)~\cite{xiu2022icon}, and neural radiance fields (NeRFs)~\cite{mildenhall2021nerf, peng2021neural, chen2021animatable, weng2022humannerf, li2022tava, li2023posevocab, isik2023humanrf} have been developed to learn the clothed human body using volume rendering techniques. However, due to the large consumption of the volumetric learning process, these methods could not balance well the training efficiency and visual quality.

Recent advances in 3D Gaussian Splatting~\cite{kerbl20233d} based methods have shown promising performances and less time consumption in this area, covering both single-view~\cite{li2024gaussianbody, kocabas2023hugs, lei2023gart, hu2023gauhuman} and multi-view~\cite{li2023animatable, zielonka2023drivable} avatar reconstruction settings. Beyond all these works, two main ongoing challenges still need to be resolved. The first one is efficiently training the Gaussian Splatting models across different poses and the second is improving the visual quality for dynamic details. 

For the dynamic pose learning problem, there are several existing works~\cite{ye2023animatable,qian20243dgsavatar} that have already adopted the pose-dependent deformation from SMPL~\cite{SMPL:2015} prior. Unfortunately, they are all limited by the global pose vectors and neural skinning weights learning and hence lack the local geometry correspondence for clothed human details. To address this limitation, our insight is to enable the point-level SMPL deformation prior to training 3D Gaussian Splatting avatar with local pose guidance. Specifically, we take inspiration from \textit{SCARF}~\cite{feng2022capturing} by deforming the avatar with SMPL-KNN strategy and \textit{Deformable-GS}~\cite{yang2023deformable3dgs} by incorporating position and deformation codes into a Multilayer Perceptron (MLP). This approach enables the learning of locally non-rigid deformations, which are subsequently transformed using rigid deformation to align the adjusted model with the observed space. In this way, our model can efficiently learn the local geometric prior information from SMPL deformation and maintain correspondence consistency for cloth details across all the frames.

For the visual quality problem, we observe that the current 3D Gaussian Splatting model is struggling to render the non-rigidly deformed human avatars in high fidelity. We decouple the visual quality issue into two parts and propose two technical solutions correspondingly. The first issue is the unstable rendering results caused by complex non-rigid deformation between different pose spaces and the canonical space. To overcome that, we optimize a physically-based prior for the Gaussians in the observation space to mitigate the risk of overfitting Gaussian parameters. We transform the local rigid loss \cite{luiten2023dynamic} to regularize over-rotation across the canonical and observation space. The second issue is that the original Gaussian Splatting sampling strategy could not well handle the rich texture details like complicated clothes. We tackle this problem by introducing a split-with-scale strategy to further enhance the geometry expressiveness of the Gaussian Splatting model and resolve the visual artifacts in texture-rich areas.

%The inherent limitations of 3D-Gaussian Splatting in accurately reconstructing human motion scenes underscore the complexity of managing model deformations between pose space and canonical space. In light of this challenge, we embark on a concerted effort to address this issue by incorporating the physics-based prior inspired by \cite{luiten2023dynamic}. Additionally, we undertake an extensive exploration of scaling methodologies, aimed at refining Gaussian splats and thereby enhancing their effectiveness in capturing intricate motion dynamics.  
\begin{figure*}
    \centering
    \includegraphics[width=1\linewidth]{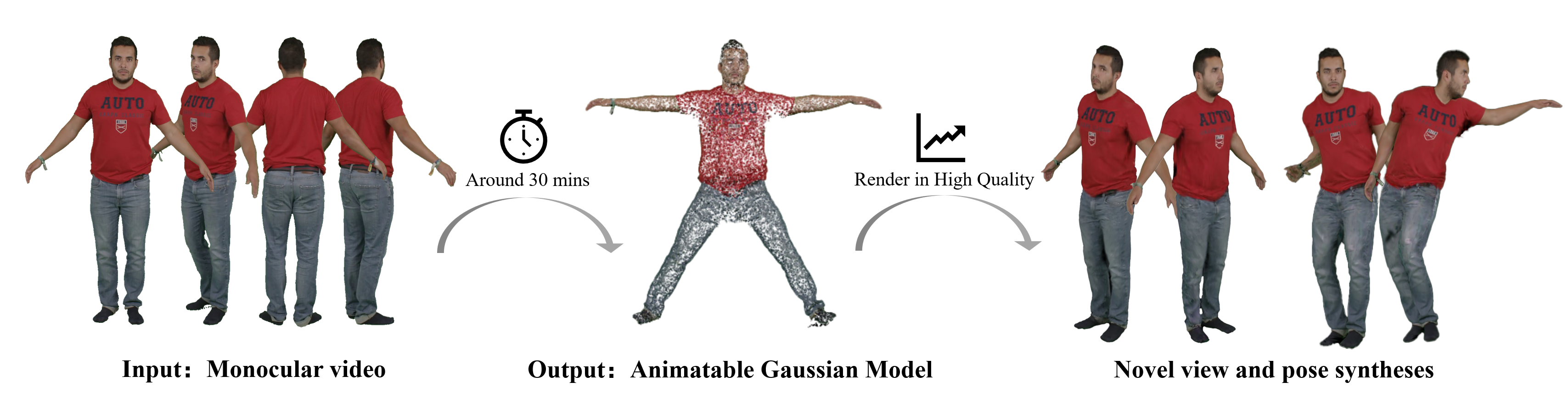}
    \caption{AniGaussian takes monocular RGB video as input, reconstructing an animatable avatar model in around 30 minutes and rendering with 45 FPS on a single NVIDIA RTX 4090 GPU. The resulting human model can present subtitle texture and generate non-rigid deformation of clothes details. Performance in novel views and animation with unseen poses. Furthermore, we gain the highest reconstruction quality in current works which is evident in our picture metrics.}
    \label{fig:first}
\end{figure*}
Based on the above analysis of the current limitations for Gaussian based animatable avatar models, we combine our insights and propose another novel framework called \textit{AniGaussian}. Our framework extends the 3D-GS representation to animatable avatar reconstruction, with an emphasis on enabling local pose-dependent guidance and visual quality refinement. Given a monocular human avatar video as input, \textit{AniGaussian} can efficiently train an animatable Gaussian model for the full-body avatar in 30 minutes as shown in Figure~\ref{fig:first}. In the experiment, we evaluate our proposed framework on monocular videos of animatable avatars on the task of novel view synthesis and novel pose synthesis. By comparing it with other works, our method achieves superior reconstruction quality in rendering details and geometry recovery, while requiring much less training time and real-time rendering speed. We conduct ablation studies to validate the effectiveness of each component in our method.

%In this work, we present the AniGaussian, which extended the 3D-GS representation to animatable avatar reconstruction by utilizing an articulated human model as guidance. Specifically, we decomposed the pose-guided deformation into non-rigid and rigid deformation. The non-rigid deformation presents the cloth's dynamic details and deforms the 3D Gaussians into motion space with the control of the corresponding SMPL. To avoid the uncertain SMPL parameter influencing the model train, we jointly optimize the SMPL parameters. Secondly, we optimize a physically-based prior for the Gaussians in the observation space to mitigate the risk of overfitting Gaussian parameters. We transform the local rigid loss \cite{luiten2023dynamic} to regularize over-rotation across the canonical and observation space and propose a split-with-scale strategy to enhance point cloud density and avoid artifacts. In the experiment, we evaluate our proposed framework on monocular videos of animatable avatars on the task of novel view synthesis and novel pose synthesis. By comparing it with other works, our method achieves superior reconstruction quality in rendering details and geometry recovery, while requiring much less training time and almost real-time rendering speed. We also conduct ablation studies to validate the effectiveness of each component in our method.

% In summary, Our work has the following contributions:
% \item We propose a method to reconstruct dynamic human motion scene with pose-guided deformation based on 3D-GS
In summary, our contributions are as follows:
\begin{itemize}
    \item A pose-guided deformation framework that includes both non-rigid and rigid deformation to extend the 3D Gausssian Splatting to animatable avatar reconstruction.
    \item We advanced Gaussian Splatting with the rigid-based prior restricting the canonical model and Split with scale strategy to achieve more accuracy and robustness.
    \item Our approach has yielded the best results on the PeopleSnapshot dataset, demonstrating superior rendering quality compared to other methods.
\end{itemize}

\begin{figure*}
    \centering
    \includegraphics[width=1\linewidth]{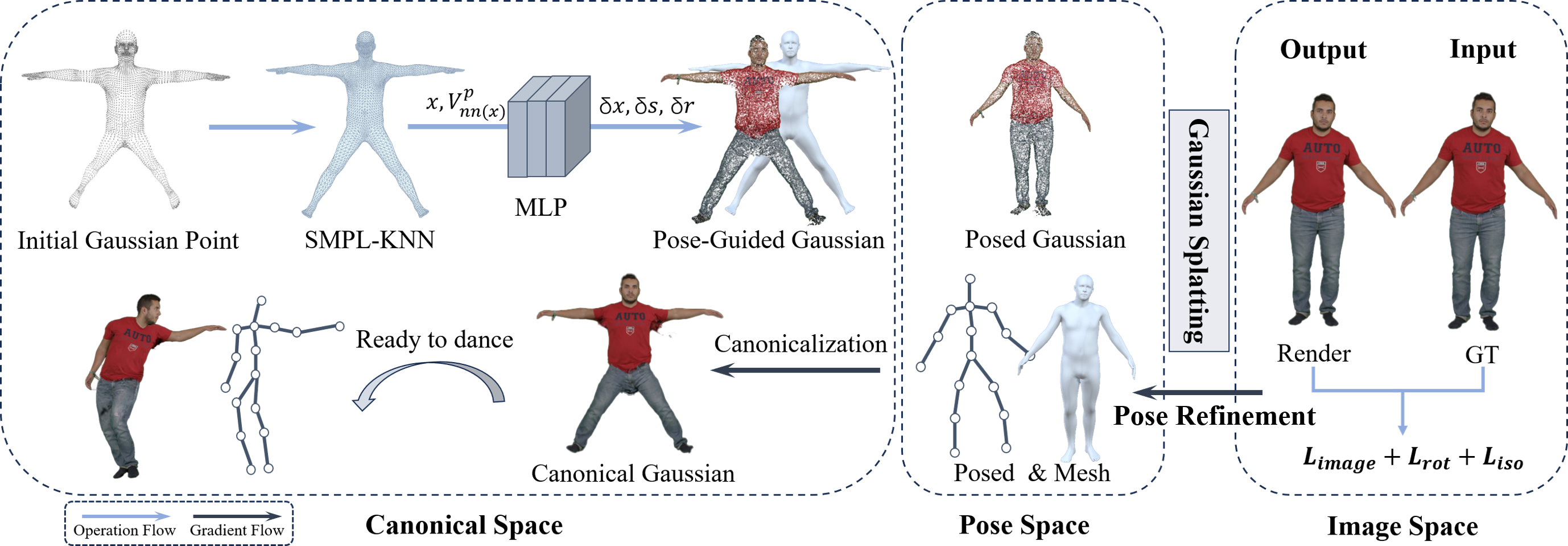}
    \caption{\textbf{Overview of AniGaussian.}
        At first, we initialize the point cloud using SMPL vertices. In the train processing, we find the nearest vertex as the deformation-guider of the Gaussian. We input the position of Gaussian after position encoding and the nearest vertex as the deformation code to the MLP to gain the non-rigid deformation. Then with the transformation of the SMPL vertex, the Gaussians are transformed to the pose space. In the tour of transformation, we use the rigid-based prior $L_{rot}$ and $L_{iso}$ to rule the deformation. After Gaussian splatting, we could refine the SMPL parameters and the canonical model.
    }
    \label{fig:pipeline}
\end{figure*}

\section{Related Work}
\label{sec:relatedwork}
\subsection{Animatable Avatar Reconstruction}
Reconstructing 3D humans from images or videos is a challenging task.
Recent works~\cite{alldieck2018detailed,alldieck2018video,ma2020learning} use morphable mesh models like SMPL~\cite{SMPL:2015} to reconstruct 3D humans from monocular videos or single images.
However, explicit mesh representations are incapable of capturing intricate clothing details. 

To address these limitations, neural representations have been introduced \cite{saito2019pifu,saito2020pifuhd,han2023high} for 3D human reconstruction. Implicit representations, like PIFU~\cite{saito2019pifu} and its variants, achieve impressive results in handling complex details such as hairstyle and clothing while . ICON~\cite{xiu2022icon} and ECON~\cite{xiu2023econ} leverage SMPL prior to handling extreme poses. Other methods \cite{zheng2021pamir,huang2020arch,he2021arch++} use parametric models to handle dynamic scenes and obtain animatable 3D human models. Recent advancements involve using neural networks for representing dynamic human models. Extensions of NeRF \cite{mildenhall2021nerf} into dynamic scenes \cite{pumarola2021d,park2021nerfies,park2021hypernerf} and methods for animatable 3D human models in multi-view scenarios \cite{isik2023humanrf,lin2023im4d,peng2021animatable,li2022tava,li2023posevocab,weng2022humannerf} or monocular videos \cite{zhao2022human,peng2021neural,chen2021animatable,jiang2023instantavatar} have shown promising results. Signal Distance Function (SDF) is also employed \cite{liao2023high,jiang2022selfrecon,guo2023vid2avatar} to establish a differentiable rendering framework or use NeRF-based volume rendering to estimate the surface. However, most implicit representations are unfortunately struggling to handle the balance between the cost of long training process and achieving high quality rendering result.

3D Gaussian Splatting (3D-GS) model~\cite{kerbl20233d} is deemed as a promising improvement of the previous implicit representations. With 3D-GS backbone, the training and inference speed could be improved by reducing a large amount of time. In this work, we incorporate the latest 3D-GS idea into the animatable avatar reconstruction topic to enhance both the time efficiency and training robustness.

\subsection{Dynamic Gaussian Splatting}
Similar to NeRF, 3D-GS could reconstruct dynamic scenes from multi-view pictures with an additional network with the time features\cite{yang2023deformable3dgs, Wu_2024_CVPR} or with rigidly physical-based prior\cite{xie2023physgaussian, feng2024gaussian, jiang2024vrgs}. With control ability of the explicit point cloud, SC-GS\cite{huang2023sc} combines 3D Gaussian with a learnable graph to provide a control layer to deform the gaussian splats and corresponding features.

Many recent works also try to model 3D-GS avatars with human body prior like SMPL. With multi-view input, Animatbale 3D Gaussian~\cite{li2023animatable} adopts the SDF representation as the geometry proxy and introduces 2D-CNNs to generate the Gaussian map as neural texture. With single-view input, GaussianAvatar~\cite{hu2023gaussianavatar} employs the UV texture of SMPL as the pose feature to generate a Gaussian point cloud. SplattingAvatar~\cite{shao2024splattingavatar} binds the Gaussian point with triangular mesh facet along with additional translation on surface. Other methods\cite{lei2023gart, hu2023gauhuman, kocabas2023hugs} use the learnable skinning weight to associate the Gaussian point cloud to the bone transformation. However, these methods do not consider the local pose-dependent deformation and thus fail to efficiently use the local guidance of SMPL prior. In this work, our method targets at learning the Gaussian Splatting models across pose-deformed frames and improves the visual quality for dynamic details.

\section{Method}
In this section, we first describe our framework pipeline for 3D-GS based animatable avatar reconstruction. Then we elaborate on pose-guided local deformation to train the dynamic Gaussian. Finally, we introduce the advanced gaussian splatting to regularize the 3D Gaussians across the canonical and observation spaces. 
\subsection{Overview}
\label{subsec:overview}
As shown in Figure. \ref{fig:pipeline}, we initialize the point cloud with the SMPL vertex in the star-pose and define the template 3D Gaussians in the canonical space as $G(\bar{x},\bar{r},\bar{s},\bar{\alpha},\bar{f})$. We decompose the animatable avatar modeling problem into the canonical space and the pose space. To learn the template 3D Gaussians, we employ pose-guidance deformation fields to transform them into the pose space and render the scene using differentiable rendering. In order to reduce the artifacts of 3D Gaussian with invalid rotations or unexpected movements in canonical space, we constrain the 3D Gaussians with the rigid-based Prior. Finally, to handle the rich texture details like complicated clothes, we further refine the naive gaussian splatting approach with a split-with-scale strategy to enhance the expressiveness of our model and resolve the visual artifacts in texture-rich areas.
\subsection{Pose-guided Deformation}
\label{subsec:pgd}
We utilize the parametric body model SMPL \cite{SMPL:2015} as pose guidance. 
The articulated SMPL model $M(\beta,\theta)$ is defined with pose parameters $\theta \in R^{69}$ and shape parameters $\beta \in R^{10}$ that outputs a 3D human body mesh with vertices $V \in R^{6890\times3}$, and vertex transform $T(\beta,\theta)$ from the T-pose.
To gain the transformation from the SMPL model, we find the nearest vertex of canonical 3D Gaussians, register it as the agent, on the template model $V^{c} = M(\beta,\theta_c)$ that in the star-pose as shown in Figure.\ref{fig:pipeline}. 

In order to fully utilize the local correspondence information provided by SMPL prior, we take inspirations from SelfRecon~\cite{jiang2022selfrecon} and SCARF~\cite{feng2022capturing} and decompose the pose-guided deformation fields into non-rigid transformation for the cloth movement and rigid transformation for the body movement. 

\noindent \textbf{Non-rigid transformation. }
First we implement a MLP $F$ to learn the non-rigid deformation of the cloth details, 
\begin{equation}
    F(x, V^{p}_{nn(x)}) = \delta x, \delta r, \delta s,
    \label{eq:non_rigid mlp}
\end{equation}
this MLP takes as input the position of the 3D Gaussian $x$ and the position of posed SMPL model vertex $V^{p}_{nn(x)}$, and output the $\delta x$, $\delta r$, $\delta s$ as gaussian parameters. 
The $V^{p}_{nn(x)}$ is the vertex on the posed SMPL model that contains the same index of the template model $V^{c}$.
And our canonical model after non-rigid deformation is $G(\bar{x}',\bar{r}',\bar{s}',\bar{\alpha},\bar{f})$.

\noindent \textbf{Rigid transformation. }
The rigid transformation from canonical space to observation space of 3D Gaussians is defined by the transformation of SMPL vertex as: 
\begin{equation}
    D(\bar{x},\beta,\theta_t,\theta_c) = \sum_{\substack{v^{c}_i \in nn(\bar{x})}}\frac{\mathbf{w}_i}{\mathbf{w}}T_i(\beta,\theta_{c})^{-1}T_i(\beta,\theta_{t}),
    \label{eq:Gaussian_rigid_deform}
\end{equation}
where $v^{c}_i$ is one of the $k$ nearest vertex of template model and $T_i$ is the transformation of the vertex. $\theta_c$ is the predefined canonical pose parameter, so we omit it in Eq.~\ref{eq:Gaussiandeform}. $\theta_t$ is the pose of current frame. We set $k=3$ to maintain the 3D Gaussians transformation stability across multiple joints, and further weigh the transformations with:
\begin{equation}
    \begin{split}
    \mathbf{w}_i(x) = exp(-&\frac{||x-vi||_2||w_{nn(x)}-w_i||_2}{2\sigma^2}),\\
    \mathbf{w}(x) &= \sum_{\substack{v^{c}_i \in nn(x)}}\mathbf{w}_i(x),
    \end{split}
    \label{eq:Gaussian_rigid_weight}
\end{equation}
where $\sigma = 0.1$, $w_{nn(x)}$ is the skinning weight of the $k$ nearest vertex, $w_i$ is the blend weight of nearest vertex.

For each frame, we transform the position $\bar{x}'$ and rotation $\bar{r}'$ of the canonical Gaussians after non-rigid deformation to the observation space, with the guided of pose parameter $\theta_t$ of current frame and the global shape parameter $\beta$:
\begin{equation}
\begin{split}
    x &= \mathcal{D}(\bar{x},\theta_t,\beta)\bar{x}',\\
    r &= \mathcal{D}(\bar{x},\theta_t,\beta)\bar{r}',
\end{split}
    \label{eq:Gaussiandeform}
\end{equation}
where $\mathcal{D}$ is the deformation function defined in Eq.\ref{eq:Gaussian_rigid_deform}. 

In this way, we obtain the deformed Gaussians in the observation space. After differentiable rendering and image loss calculation, the gradients will be passed through the inverse of the deformation field $\mathcal{D}$ and optimized parameters of the Gaussians in canonical space.

Additionally, because the monocular input is hard to provide sufficient view information, it is noteworthy to mention that we opt to transform the direction of light into the canonical space to ensure the view consistent. The light direction transformation can be formulated as:
\begin{equation}
    \bar{d} = (T_{c2w} r)^T d,
    \label{eq:lighttransf}
\end{equation}
where $d$ is the light direction in the world coordinate system, $r$ is the rotation in camera coordinate system, and $T_{c2w}$ is the coordinate transformation matrix from the camera to the world coordinate system. At last we evaluate the spherical harmonics coefficients with the canonical light direction $\bar{d}$.

\noindent \textbf{Joint optimization of SMPL parameters. }
Since our 3D-GS training pipeline is built upon the local pose-dependent deformation from SMPL prior, it is crucial to obtain accurate SMPL shapes to guarantee the pose guidance effectiveness. Unfortunately, the regression of SMPL parameters from images would be affected by a lot of reasons like false landmark detection or uncertain camera pose estimation. 

Therefore, we propose a joint optimization idea for refining the SMPL parameters including the pose and shape during training our entire pipeline. Specifically, the SMPL shape parameter $\beta$ and pose parameters $\theta$ would be optimized regarding the image loss and get updated to match the exact body shapes and poses in training frames.
%, there is a risk of generating a s. The inaccurate SMPL parameters may impact the non-rigid deformation to align with the images, leading to blurred textures. To address this issue, we propose an optimization approach for the SMPL parameters. Specifically, we designate the SMPL shape parameter $\beta$ and pose parameters $\theta$ as the optimized parameters and refine them through the optimization process. As part of pose-guided deformation, they could be guided by defined losses.

\subsection{Advance Gaussian Splatting}
\label{subsec:PBP}
Since we define the Gaussians in the canonical space and deform them to the observation space for differentiable rendering, the optimization process is still an ill-posed problem. Because multiple canonical positions will be mapped to the same observation position, there are inevitably overfitting in the observation space and visual artifacts in the canonical space. To address this problem, we propose an advanced gaussian splatting to enhance the visual performance.

\begin{figure}
        \centering
        \includegraphics[width=1\linewidth]{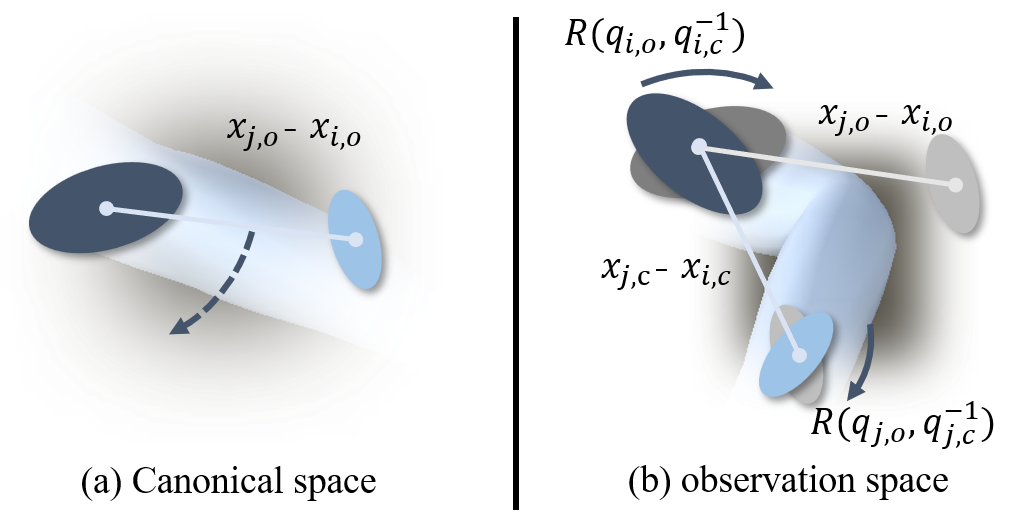}
        \caption{\textbf{Visual of the Rigid-based prior.} With the deformation between the canonical space and the observation space, we hope the neighbour Gaussian could have a similar rotation and keep a property distance.}
        \label{fig:physical-based}
\end{figure}

\noindent \textbf{Rigid-based prior. }
In the experiment, we also observed that this optimization approach might easily result in the novel view synthesis showcasing numerous Gaussians in incorrect rotations, consequently generating unexpected glitches.
Thus we follow \cite{luiten2023dynamic} to regularize the movement of 3D Gaussians by their local information. Particularly we employ two regularization losses to maintain the local geometry property of the deformed 3D Gaussians, including local-rotation loss $\mathcal{L}_{rot}$ and a local-isometry loss $\mathcal{L}_{iso}$.
Different from \cite{luiten2023dynamic} that attempts to track the Gaussians frame by frame, we regularize the Gaussian transformation from the canonical space to the observation space. And we do not set the rigid loss because of it would conflict with the non-rigid deformations.

Given the set of Gaussians $j$ with the k-nearest-neighbors of $i$ in canonical space (k=5), the isotropic weighting factor between the nearby Gaussians is calculated as:
\begin{equation}
    w_{i,j} = exp(-\lambda_{w}||x_{j,c}-x_{i,c}||^{2}_{2}),
    \label{eq:local-weight}
\end{equation}
where $||x_{j,c}-x_{i,c}||$ is the distance between the Gasussians $i$ and $j$ in canonical space, set $\lambda_{w} = 2000$ that gives a standard deviation. The rotation loss could enhance convergence to explicitly enforce identical rotations among neighboring Gaussians in both spaces:
\begin{equation}
    \mathcal{L}_{rot} =  \frac{1}{k|{G}|}\sum_{i \in {G}}\sum_{j \in knn_{i;k}}w_{i,j}||q_{j,o}q^{-1}_{j,c}-q_{i,o}q^{-1}_{i,c}||_{2},
    \label{eq:local-rot}
\end{equation}
where $G$ is the whole Gaussian model, $q$ is the normalized Quaternion representation of each Gaussian's rotation, the $q_{o}q^{-1}_{c}$ demonstrates the rotation of the Gaussians from the canonical space to the observation space. The $w_{i,j}$ is the weighting factor as mentioned in Eq.~\ref{eq:local-weight}.

We use an isometric constraint to make two Gaussians in different spaces in a property distance to avoid floating artifacts, which enforces the distances $\Delta x = x_i-x_j$ in different spaces between their neighbors:
\begin{equation}
    \mathcal{L}_{iso}=\frac{1}{k|{G}|}\sum_{i \in {G}}\sum_{j \in knn_{i;k}}w_{i,j}\{||\Delta x_o||_{2}-||\Delta x_c||_{2}\},
    \label{eq:iso-rigidity}
\end{equation}
after adding the above objectives, our objective is :
\begin{equation}
    \mathcal{L}=\mathcal{L}_{L1}+\lambda_{SSIM}\mathcal{L}_{SSIM}+\lambda_{rot}\mathcal{L}_{rot}+\lambda_{iso}\mathcal{L}_{iso}.
    \label{eq:losses}
\end{equation}
where $\mathcal{L}_{L1}$ and $\mathcal{L}_{SSIM}$ are the images losses from original 3D-GS~\cite{kerbl20233d}, which regular the model from the image space to optimize the Gaussians and the other models in our method, the $\lambda_{SSIM}$, $\lambda_{rot}$ and $\lambda_{iso}$ are loss weight.

\noindent\textbf{Split-with-scale. }
After adjusting to utilize monocular video input, the model lacks some of the geometric information obtained from multi-view sources. A portion of the reconstructed point cloud (3D Gaussians) may become excessively sparse, leading to oversized Gaussians and to generate blurring artifacts with novel motions. To address this, we propose a strategy to split large Gaussians using a scale threshold $\epsilon_{scale}$ after the regular split and densify. If a Gaussian has scale $s$ larger than $\epsilon_{scale}$, we decompose it into two identical Gaussians, each with half the size. With such operation, we could gain a more compact Gaussian model. The compact Gaussian would preserve more geometry information to avoid confusion by the texture.

\noindent\textbf{Initial with SMPL vertex. }
For the reconstruction of the 3D Gaussian model, point clouds are required as the basis input. The original 3D-GS used COLMAP to initialize multi-view images to generate the basic point clouds. However, for monocular image input, it is not possible to use COLMAP to generate the basic point clouds. But based on prior knowledge of the human body, we can use the vertices of human mesh as the basic point clouds for the reconstruction.
\begin{table*}
\centering
\resizebox{\textwidth}{!}{   
    \begin{tabular}{lcccccccccccccc}
        \toprule
                    &&&\multicolumn{3}{c}{male-3-casual} & \multicolumn{3}{c}{male-4-casual} & \multicolumn{3}{c}{female-3-casual} & \multicolumn{3}{c}{female-4-casual}\\
                    &\textbf{time↓}&\textbf{FPS↑}&\textbf{PSNR↑} & \textbf{SSIM↑} & \textbf{LPIPS↓}  & \textbf{PSNR↑} & \textbf{SSIM↑} & \textbf{LPIPS↓} & \textbf{PSNR↑} & \textbf{SSIM↑} & \textbf{LPIPS↓} & \textbf{PSNR↑} & \textbf{SSIM↑} & \textbf{LPIPS↓} \\
        \midrule
             3D-GS\hfill        & 0.5h& 70&  \colorbox{lightgray}{26.60} &\colorbox{lightgray}{0.9393} &\colorbox{lightgray}{0.082} &\colorbox{lightgray}{24.54} & \colorbox{lightgray}{0.9469} &\colorbox{lightgray}{0.088} &\colorbox{lightgray}{24.73} &\colorbox{lightgray}{0.9297} & \colorbox{lightgray}{0.093} &\colorbox{lightgray}{25.74} &\colorbox{lightgray}{0.9364} &\colorbox{lightgray}{0.075} \\
             Anim-NeRF\hfill      &26h& 1 & 29.37 &0.9703 &0.017 &28.37 &0.9605 & 0.027 &28.91 & {0.9743}&  {0.022} &28.90 &0.9678 &0.017 \\
             InstantAvatar\hfill&\underline{0.1h}& 15  & 29.64 &0.9719 &0.019 &28.03 &0.9647 &0.038 &28.27&0.9723& 0.025&29.58 &0.9713 &0.020 \\
             GauHuman*\hfill &\underline{0.1h}& \underline{300} &30.95 &0.9504&0.046&28.28 &0.9523&0.055& {32.52}&0.9627&0.058&30.02 &0.9494&0.044 \\
             GART\hfil     &\underline{0.1h}&90& 30.40 & \colorbox{pink}{0.9769} &0.037 &27.57 & {0.9657} &0.060 &26.26 &0.9656 &0.049 &29.23 & {0.9720} &0.037 \\
        \midrule
             3DGS-Avatar\hfill &0.75h&\underline{50}& {34.28}& 0.9724 & {0.014} & {30.22} &0.9653 & {0.023} &30.57& 0.9581 &0.020 & \colorbox{pink}{33.16}& 0.9678 &0.016  \\
             \textbf{Ours}  &\underline{0.5h}&45& \colorbox{pink}{35.35}& {0.9762}& \colorbox{pink}{0.012} & \colorbox{pink}{33.35} & \colorbox{pink}{0.9765}& \colorbox{pink}{0.018} & \colorbox{pink}{35.01}& \colorbox{pink}{0.9752}& \colorbox{pink}{0.017}& {33.02}& \colorbox{pink}{0.9798}& \colorbox{pink}{0.014} \\
        \bottomrule
    \end{tabular}}
    \caption{\textbf{Quantitative comparison of novel view synthesis on PeopleSnapshot dataset.} 
    Our approach exhibits a significant advantage in metric comparisons, showing substantial improvements in all metrics due to its superior restoration of image details.\colorbox{pink}{NUM}= The Best, \colorbox{lightgray}{NUM} = The Worst. ``*'' denotes the results trained by the official codes.}
    \label{table:compare}
\end{table*}
\begin{figure*}
        \centering
        \includegraphics[width=1\linewidth]{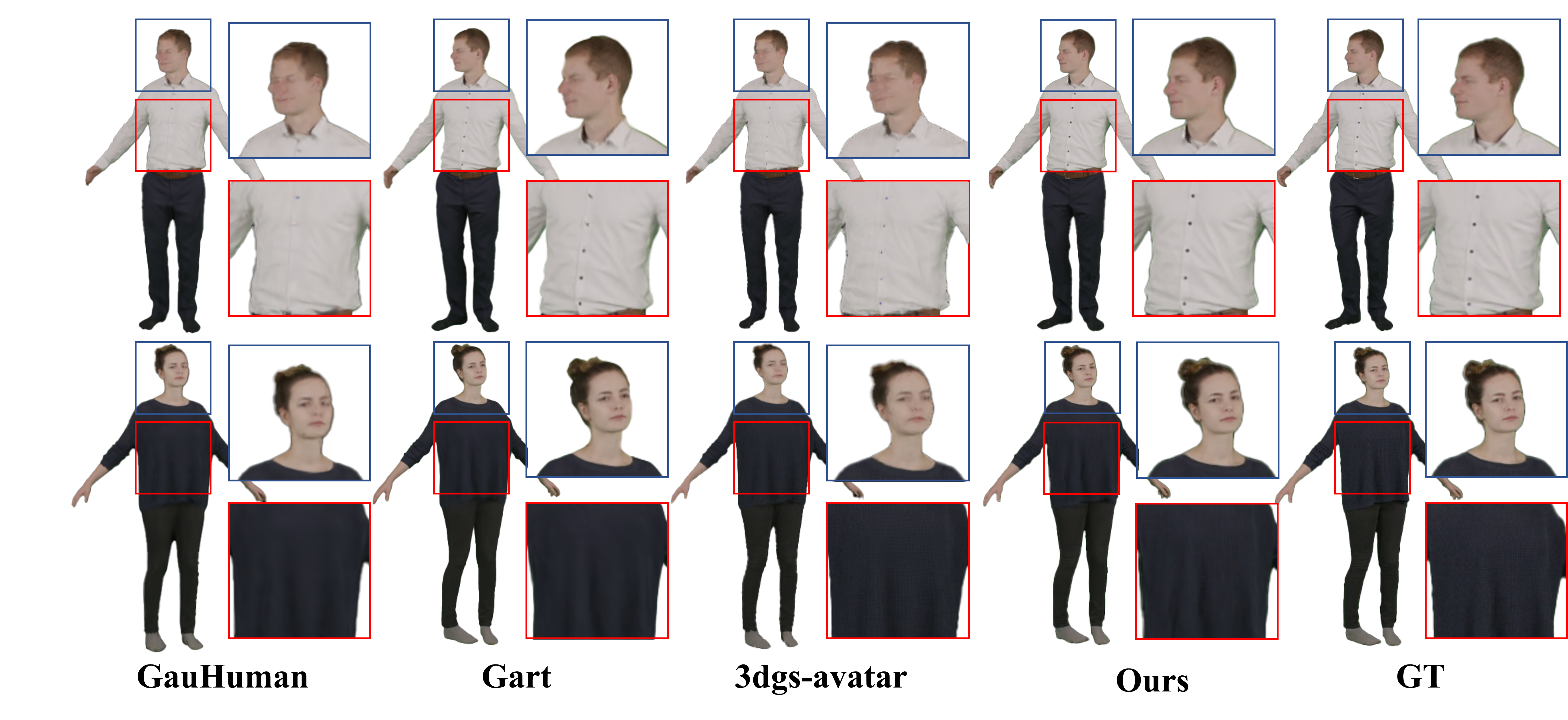}
        \caption{\textbf{Qualitative comparison of novel view synthesis on PeopleSnapshot dataset.} Compare to other methods, our method effectively restores details on the animatable avatar, including intricate details in the hair and folds in the clothes. These results underscore the applicability and robustness in real-world scenarios. }
        \label{fig:ps_compare}
\end{figure*}
% \begin{figure*}
%     \centering
%     \includegraphics[width=1\linewidth]{pic_draft/pai.png}
%     \caption{\textbf{Visual comparison of different methods about novel view synthesis on ZJU-MoCap\cite{peng2021neural}}. Our method achieves high fidelity results, especially in the texture of the clothes and the wrinkles in the garment. Compared with other methods, we have preserved more high-frequency detail from the pictures.}
%     \label{fig:zjuvisualcomapre}
% \end{figure*}
\begin{figure*}
    \centering
    \includegraphics[width=1\linewidth]{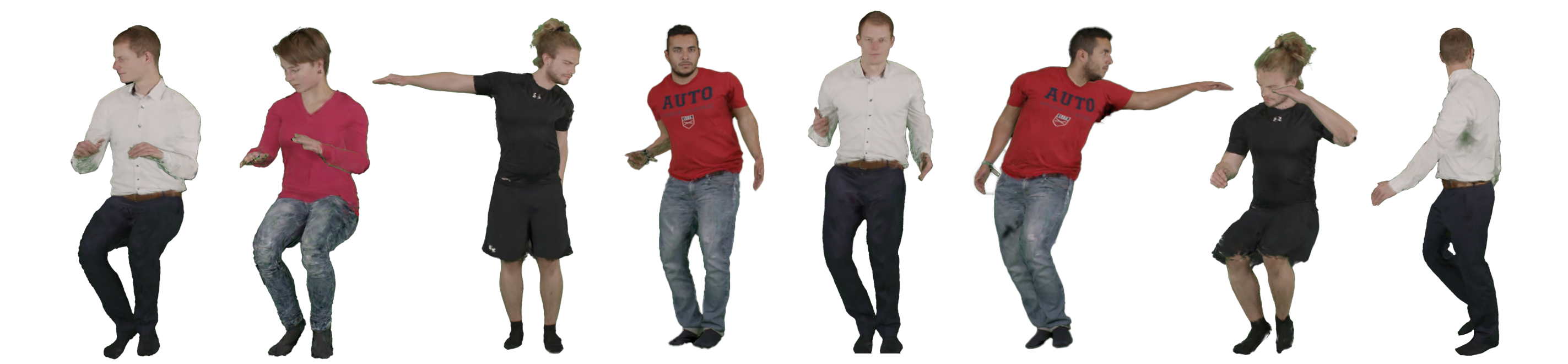}
    \caption{\textbf{Novel pose synthesis on PeopleSnapshot~\cite{alldieck2018video}}. Our method could drive the reconstruction animatable avatar in novel poses with fewer artifacts and present cloth details and render in 45FPS. 
    }
    \label{fig:novalpose}
\end{figure*}

\begin{figure*}
    \centering
    \includegraphics[width=1\linewidth]{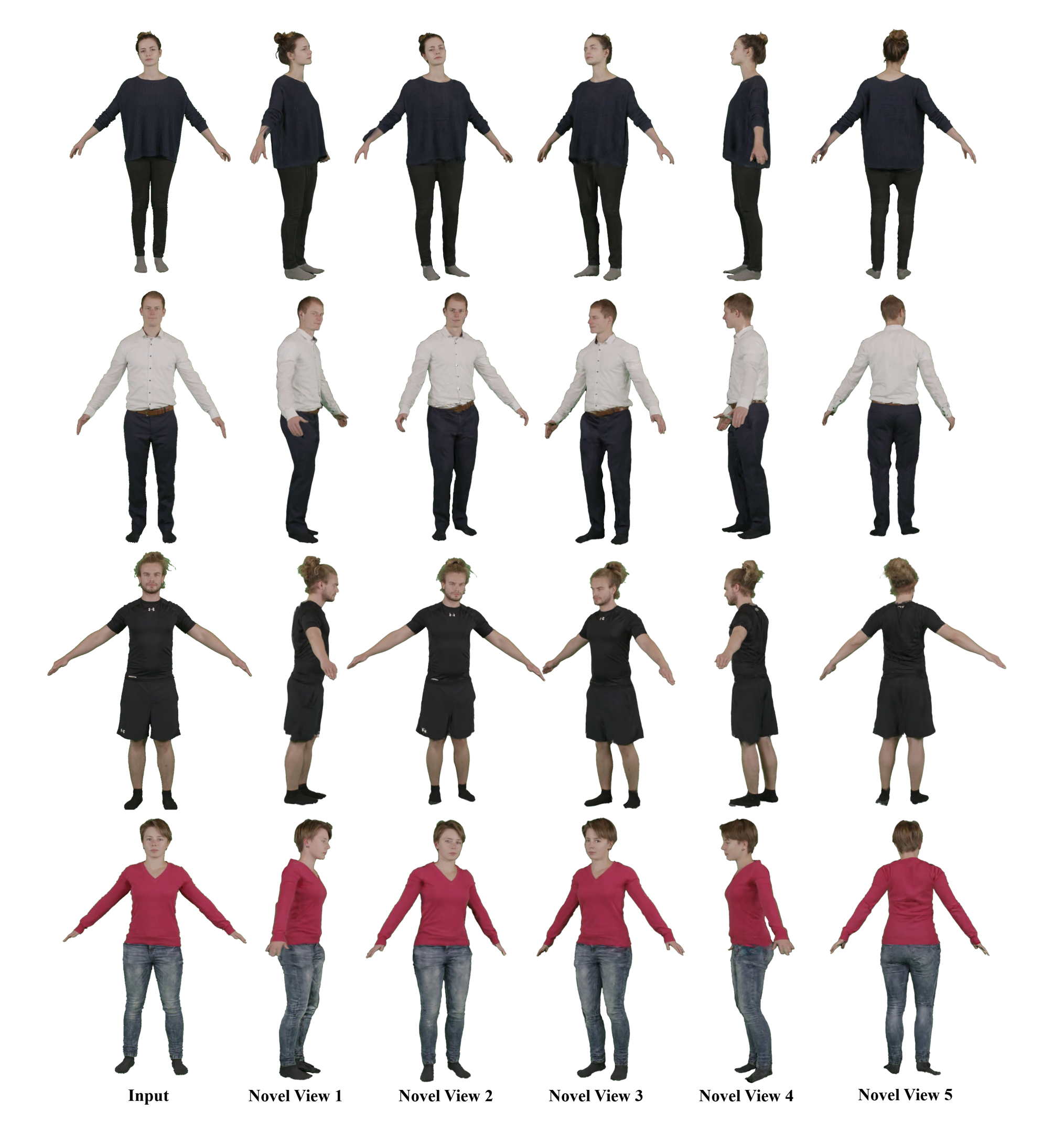}
    \caption{\textbf{Results of novel views synthesis on PeopleSnapshot \cite{alldieck2018video} dataset.} Our method effectively gene restores details on the human body, including intricate details in the hair and folds on the clothes. Moreover, the model exhibits strong consistency across different viewpoints. }
    \label{fig:arps}
\end{figure*}

\begin{figure*}
    \centering
    \includegraphics[width=1\linewidth]{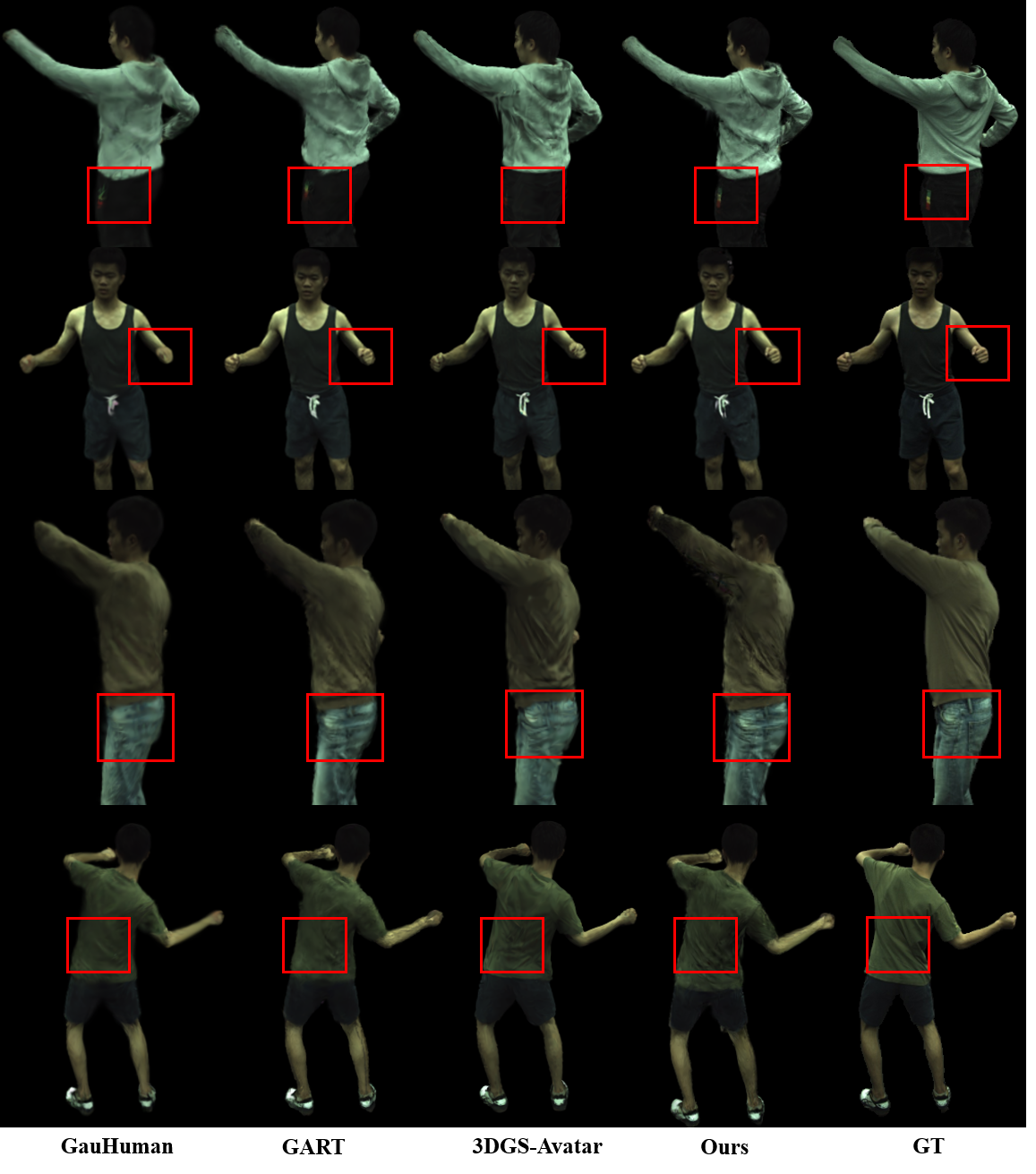}
    \caption{\textbf{Visual comparison of different methods about novel view synthesis on ZJU-MoCap\cite{peng2021neural}}. Our method achieves high fidelity results, especially in the texture of the clothes and the wrinkles in the garment. Compared with other methods, we have preserved more high-frequency detail from the pictures.}
    \label{fig:arzju}
\end{figure*}
\begin{table}
\centering
    \begin{tabular}{lccc}
    \hline
             & \textbf{PSNR↑} & \textbf{SSIM↑} & \textbf{LPIPS↓} \\ 
             \hline
    3DGS-Avatar\hfill &30.61& 0.9703 &29.58    \\
    GauHuman\hfill &31.34& 0.9647&30.51    \\
    GART\hfill & \textbf{32.22} &\textbf{0.9773} & \textbf{29.21}    \\
    \textbf{ours} & 30.00 & 0.9597 & 35.06   \\
            \hline  
    \end{tabular}
    \caption{\textbf{Metrics of novel view synthesis on ZJU-MoCap.} }
    \label{table:zju_metric}
\end{table}
\begin{table}
\centering
    \begin{tabular}{lccc}
    \hline
             & \textbf{PSNR↑} & \textbf{SSIM↑} & \textbf{LPIPS↓} \\ 
        \hline		
    Full-model &\textbf{34.18}&\textbf{0.9769}&\textbf{0.015}    \\
    w/o SMPL refine &32.31 &0.9724 &0.027   \\
    w/o $\mathcal{L}_{iso}$ &33.86 &0.9753 &0.021\\
    w/o $\mathcal{L}_{rot}$ &33.13 &0.9767 & 0.022    \\
    w/o split with scale &33.46 &0.9683 &0.020     \\
        \hline
    \end{tabular}
    \caption{\textbf{Metrics of ablation study on PeopleSnapshot.} We could gain the best picture rendering quality which has more details and is more evident in the quality of our full model.}
    \label{table:ablition study}
\end{table}
\begin{figure}
   \centering
    \includegraphics[width=1\linewidth]{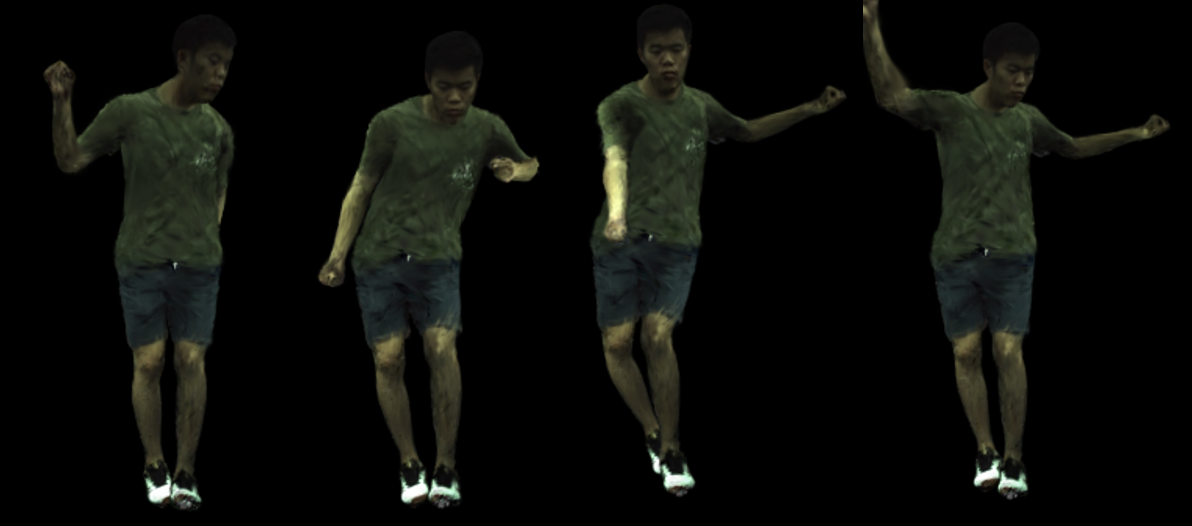}
    \captionof{figure}{\textbf{Novel poses of ZJU-MoCap~\cite{peng2021neural}.}}
  \label{fig:ad_np}
\end{figure}
\section{Experiment}
\label{sec:exp}
In this section, we evaluate our method on monocular training videos and compare the novel view synthesized result with the other benchmark. We also conduct ablation studies to verify the effectiveness of each component in our method.

% \subsection{Setup}
% %
% \noindent\textbf{Dataset and Benchmark.}
% We evaluate our method and compare with other methods on the PeopleSnapshot Dataset~\cite{alldieck2018video} and ZJU-MoCap Dataset~\cite{peng2023implicit}. Compare to the Nerf based model: InstantAvatar~\cite{jiang2023instantavatar} and Anim-NeRF~\cite{chen2021animatable}, and the Gaussian based model: 3DGS-Avatar~\cite{qian20243dgsavatar}, Gart~\cite{lei2023gart} and GauHuman~\cite{hu2023gauhuman}. The metrics settings are same as Anim-NeRF~\cite{chen2021animatable}.

\noindent \textbf{PeopleSnapshot Dataset}~\cite{alldieck2018video}  contains eight sequences of dynamic humans wearing different outfits.
The actors rotate in front of a fixed camera, maintaining an A-pose during the recording in an environment filled with stable and uniform light.
The dataset provides the shape and the pose of the human model estimate from the images.
We train the model with the frames split from Anim-nerf~\cite{chen2021animatable} and use the poses after refinement.

\noindent \textbf{ZJU-MoCap Dataset}~\cite{peng2023implicit} contains several multi-view video captures around the motion humans. This dataset has various motions and complex cloth deform. We pick 6 sequences (377, 386, 387, 392, 393, 394) from the ZJU-MoCap dataset and follow the training/test split of 3DGS-Avatar~\cite{qian20243dgsavatar}. We train the model with 100 frames captured from a stable camera view and test results with other views to measure the metrics of novel view synthesis.

\noindent \textbf{Benchmark.} On the PeopleSnapshot dataset, we compare the metrics of novel view synthesis with the original 3D-GS~\cite{kerbl20233d}, the Nerf based model: InstantAvatar~\cite{jiang2023instantavatar} and Anim-NeRF~\cite{chen2021animatable}, and the Gaussian based model: 3DGS-Avatar~\cite{qian20243dgsavatar}, Gart~\cite{lei2023gart} and GauHuman~\cite{hu2023gauhuman}. To evaluate the quality of the novel view synthesis on ZJU-MoCap, we compare it with the Gaussian-based model~\cite{qian20243dgsavatar,lei2023gart,hu2023gauhuman} on the qualitative and quantitative results.

\noindent\textbf{Performance Metrics.}
We evaluate the novel view synthesis quality with frame size in $540\times 540$ with the quantitative metrics including Peak Signal-to-Noise Ratio(PSNR)~\cite{sara2019image}, Structural SIMilarity index (SSIM)~\cite{wang2004image}, and Learned Perceptual Image Patch Similarity (LPIPS)~\cite{zhang2018unreasonable}. These metrics serve as indicators of the reconstruction quality. PSNR primarily gauges picture quality, where higher PSNR values signify demonstration illustrates the clarity of the images. SSIM measures the similarity between the ground truth and the reconstructed result, serving as an indicator of accuracy of reconstruct result. LPIPS primarily evaluates the perception of perceptual image distortion. Lower LPIPS values imply a more realistic generated images, reflecting the fidelity of the reconstruction.

\noindent\textbf{Implementation Details.}
AniGaussian is implemented in PyTorch and optimized with the Adam~\cite{2014Adam}. 
We optimize the full model in 23k steps following the learning rate setting of official implementation, while the learning rate of non-rigid deformation MLP and SMPL parameters is $2e^{-3}$. We set the hyper-parameters as $\lambda_{rot}=1$, $\lambda_{iso}=1$ and follow the original setting from 3D-GS.

\begin{figure}[htb]
   \centering
    \includegraphics[width=1\linewidth]{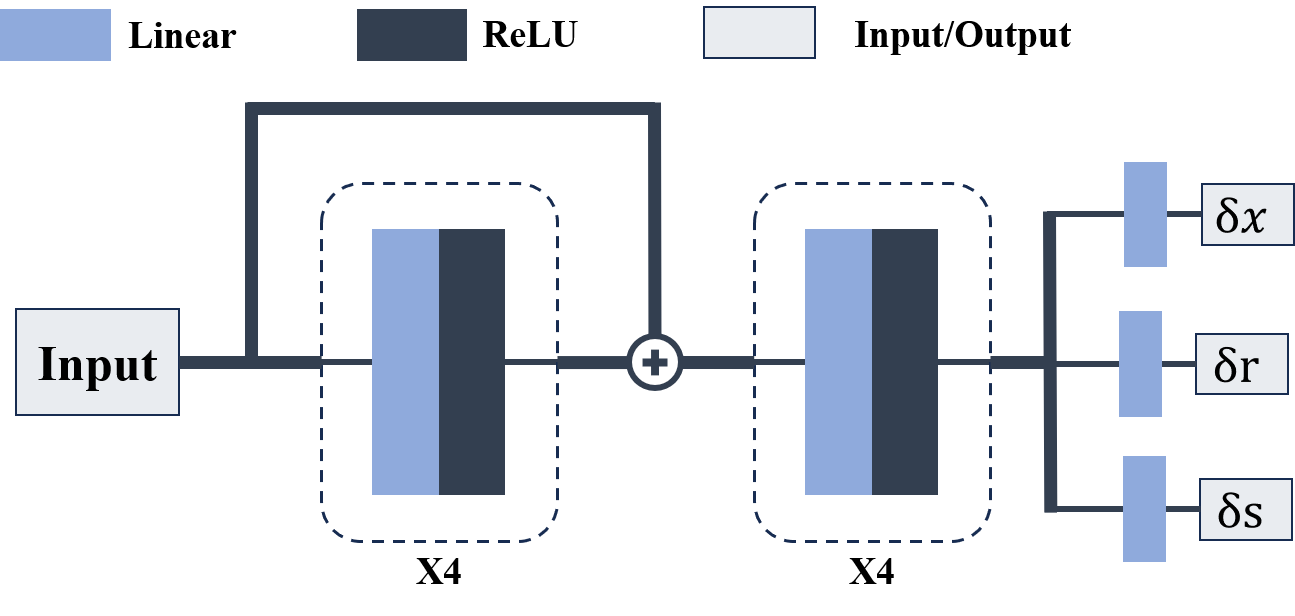}
    \captionof{figure}{\textbf{Architecture of Non-rigid Deformation Network.}}
  \label{fig:netArch}
\end{figure}

 \noindent \textbf{Non-rigid deformation network. }
 We describe the network architecture of our non-rigid deformation network in Fig.\ref{fig:netArch}. We use an MLP with 8 hidden layers of 256 dimensions which takes $x_c \in R^3 $ and deformation codes $V^{p}_{nn(x)}$ with positional encoding . Our MLP $F$ initially processes the input through eight fully connected layers that employ ReLU activations, and outputs a 256-dimensional feature vector. This vector is subsequently passed through three additional fully connected layers to separately output the offsets of position, rotation, and scaling for different pose.It should be noted that similar to NeRF~\cite{mildenhall2021nerf}, we concatenate the feature vector and the input in the fourth layer.

\subsection{Results of Novel View Synthesis}
\noindent \textbf{Quantitative analysis.}
As shown in Table~\ref{table:compare}, our method consistently outperforms other approaches in almost all metrics in the PeopleSnapshot~\cite{alldieck2018video} dataset, highlighting its superior performance in capturing detailed reconstructions. This is because our method presents the texture with high-order spherical harmonic functions and gains more accurate features by learning from the local geometry information from pose-guided deformation. Also the split-with-scale strategy benefits our model to capture more details on some challenging cases. This indicates the our model's superior performance in reconstructing intricate cloth textures and human body details. The NeRF-based methods \cite{chen2021animatable, jiang2023instantavatar} imposed by the volume rendering hardly to achieving higher quality. original 3D-GS~\cite{kerbl20233d} struggles with dynamic scenes due to violations of multi-view consistency, resulting in partial and blurred reconstructions. The test set contains variations in both viewpoints and poses, Gauhuman~\cite{hu2023gauhuman} as a method primarily focused on generating novel view synthesis, exhibits significant distortions. The deficiency in details within Gart~\cite{lei2023gart} significantly exacerbates the sense of unreality by reflecting on the higher LPIPS. We have better performance with 3DGS-Avatar~\cite{qian20243dgsavatar} in a similar training time.

In Table \ref{table:zju_metric}, AniGaussian gains comparable performance with other competitive approaches. Because our method are intentionally designed to capture the high fidelity image features and adopts a high-dimensional spherical harmonic function, which is pretty sensitive to local lighting change. However, ZJU-MoCap~\cite{peng2021neural} dataset unfortunately were not captured in a stable lighting environment. After transforming the the light directions to canonical space, the unstable lighting change would affect the training stability and thus produce some mismatching artifacts with the groundtruth. Even though, we show the rendering results of our method still demonstrate much more clear/sharp details. We argue that the quantitative metrics might not be able to reflect the model's visual quality, the compasion figure could be found in Supplementary material.

\noindent \textbf{Qualitative analysis.} 
In the comparative analysis presented in Figure~\ref{fig:ps_compare}, our method demonstrates superior performance in faithfully restoring intricate clothing details and capturing high-frequency information on the body. Unlike Gart~\cite{lei2023gart} and GauHuman~\cite{hu2023gauhuman}, which struggle to accurately reproduce texture mappings, resulting in blurry outputs almost without representation of clothing wrinkles and details, our approach excels in preserving these fine-grained features. Additionally, while 3DGS-Avatar~\cite{qian20243dgsavatar} manages to generate enough texture details, it falls short in providing high-frequency information to enhance the realism of the avatar. 

As shown in Figure \ref{fig:arps}, realistic rendering results from different views, featuring individuals with diverse clothing and hairstyles. These results underscore the applicability and robustness of our method in real-world scenarios. Additionally, it features clear and comprehensive textures, showcasing the details of both the clothing and the human body. Our method could support different types of clothes in the free-view render while maintaining a strong consistency in viewpoints.
% More Qualitative results can be found in the Supplementary material.

\subsection{Results of Novel Pose Synthesis}
\noindent \textbf{Qualitative analysis.} 
We provide the rendered result of the novel pose with the trained model in Figure~\ref{fig:novalpose}. Our reconstructed animatble avatar could perform in out-of-distribute poses that prove high-fidelity texture, such as the button on the shirt and the highlight on the belt. Artifacts in joint transformations are scarcely observed, and the reconstruction process can effectively accommodate loose-fitting garments, such as loose shorts. Benefiting from the non-rigid deformation, the complex cloth details could be preserve.

As shown in the Fig.\ref{fig:arzju}, our method demonstrates superior performance in faithfully restoring intricate clothing details and capturing high-frequency information on the body. What's more, we are also capable of generating realistic new pose effects on this dataset in Fig.\ref{fig:ad_np}.

In the comparative analysis presented in Figure~\ref{fig:arzju}, our method demonstrates superior performance in faithfully restoring intricate clothing details and capturing high-frequency information on the body. Unlike Gart~\cite{lei2023gart} and GauHuman~\cite{hu2023gauhuman}, which struggle to accurately reproduce texture mappings, resulting in blurry outputs almost without representation of clothing wrinkles and details, our approach excels in preserving these fine-grained features. Additionally, while 3DGS-Avatar~\cite{qian20243dgsavatar} manages to generate enough texture details, it falls short in providing additional high-frequency information to further enhance the realism of the avatar.
\begin{figure}[htb]
   \centering
    \includegraphics[width=0.8\linewidth]{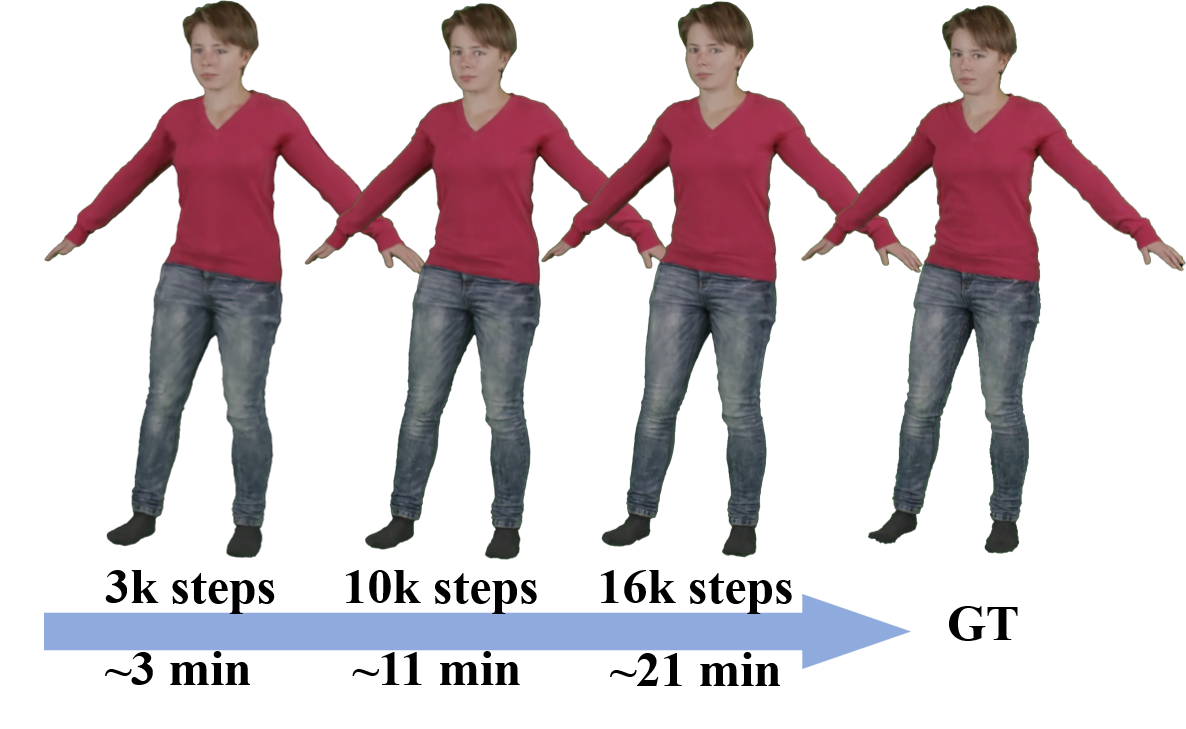}
    \captionof{figure}{\textbf{Initialization with SMPL and Efficient Reconstruct} Benefiting from the initial SMPL vertices, we can reconstruct the model in a short time. After reconstructing the basic model, our method focuses on non-rigid transformations and the details of the model.}
  \label{fig:time}
\end{figure}
\begin{figure}
    \centering
    \includegraphics[width=1\linewidth]{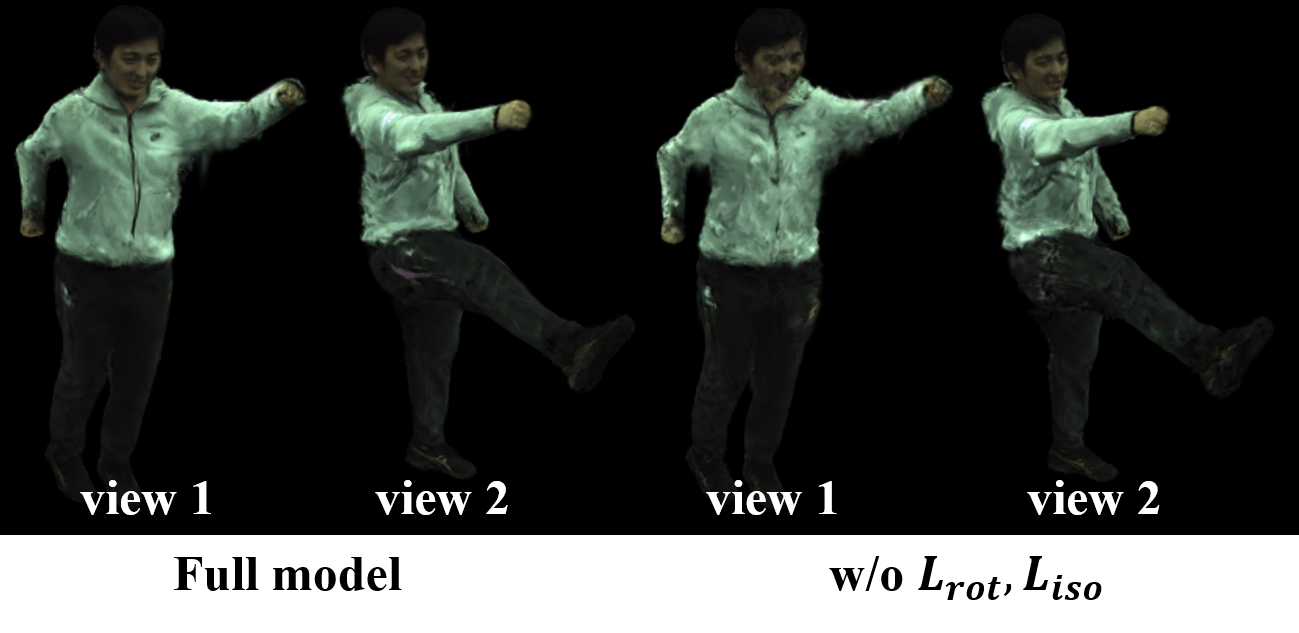}
    \caption{\textbf{Effect of rigid-based prior.} The distortions occurring under changes in viewpoint or motion. The $\mathcal{L}_{rot}$ restricts the Gaussian motion between different observation spaces, and the $\mathcal{L}_{iso}$ reduces the unexpected floating artifact. }
    \label{fig:rigid-loss}
\end{figure}
\begin{figure}
        \centering
        \includegraphics[width=1\linewidth]{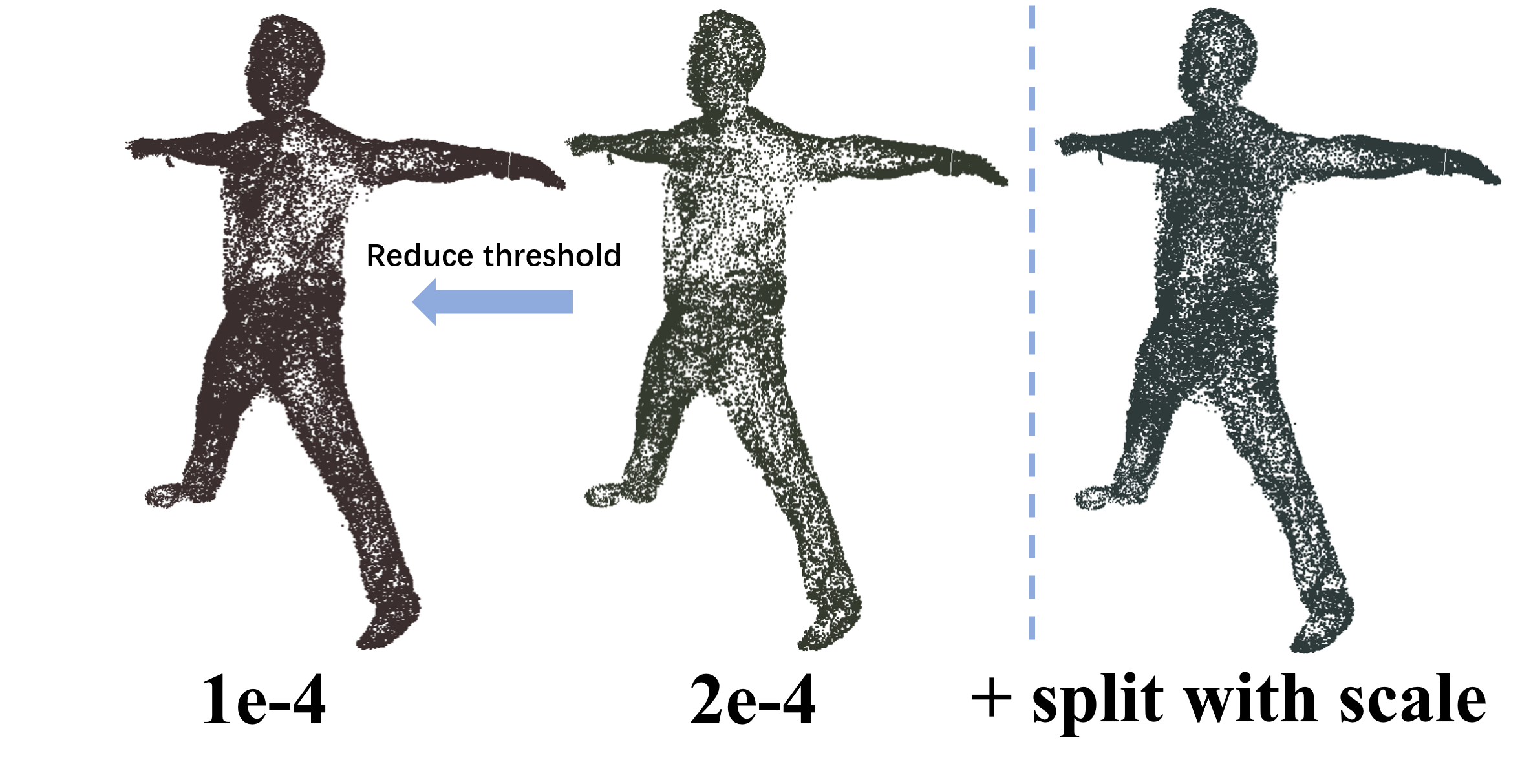}
        \caption{\textbf{Compare to the original 3D-GS split strategy} The original approach enhances geometric details by reducing the gradient threshold. The visualization of the point cloud demonstrates that our method generates denser and smoother results while effectively preventing the incorporation of texture information into the geometric domain.}
        \label{fig:split_grad}
\end{figure}
\begin{figure}
    \centering
    \includegraphics[width=1\linewidth]{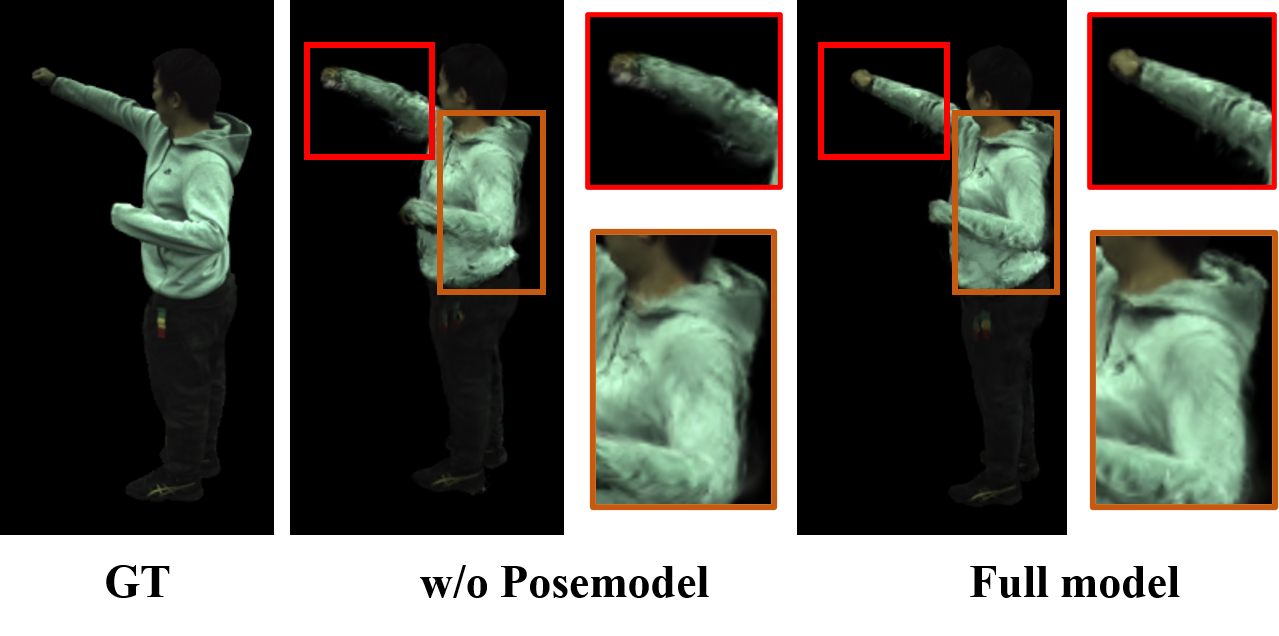}
    \caption{\textbf{Effect of pose refinement.} We compared the novel view synthesis results. The results without pose refinement would lead to floating artifacts and inexact texture. Training with the joint optimization would decrease the artifact on the sleeve and the blur texture on the collar.}
    \label{fig:refine}
\end{figure}
\begin{figure}
   \centering
    \includegraphics[width=0.8\linewidth]{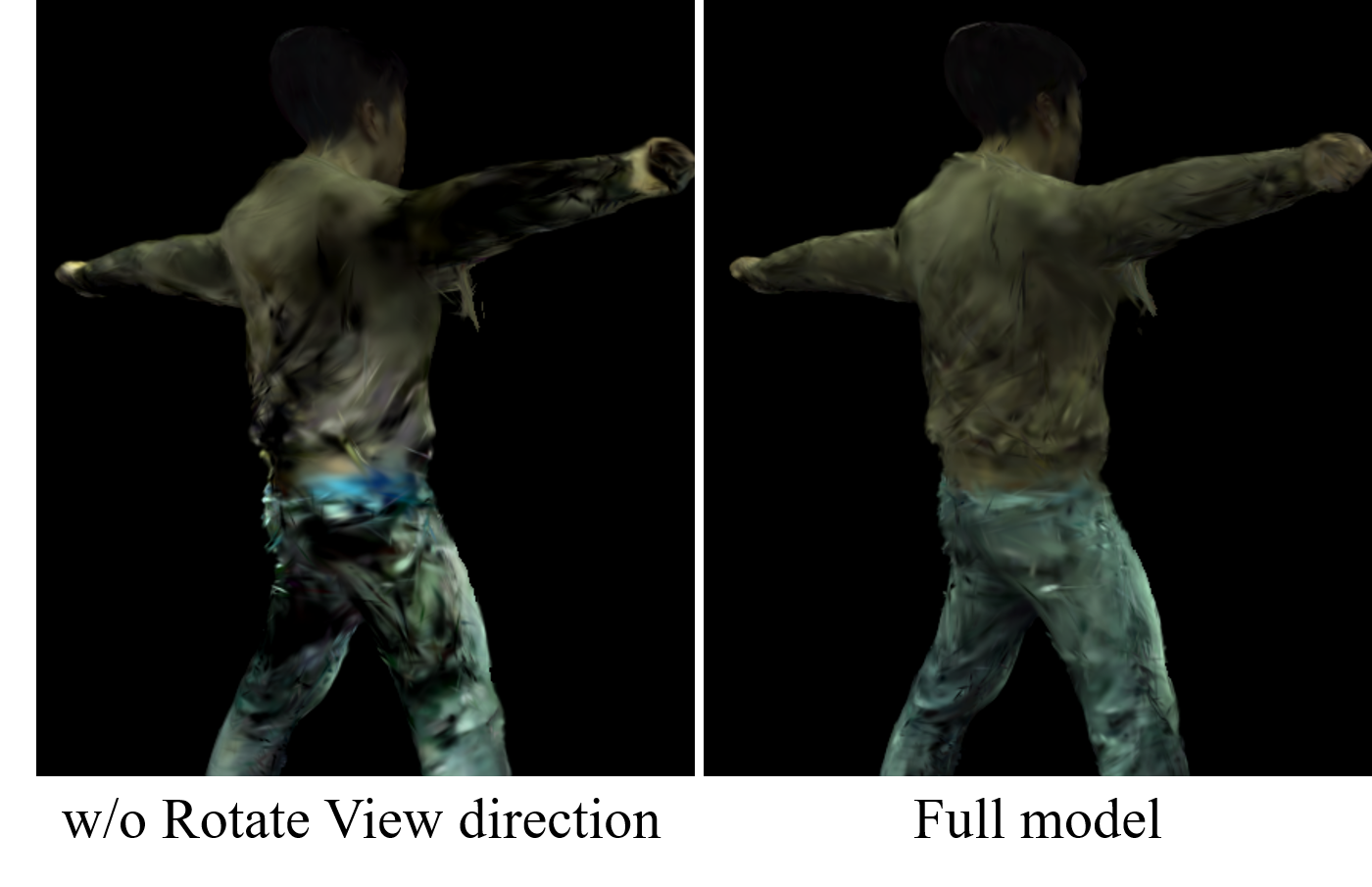}
    \captionof{figure}{\textbf{Effect of Rotating the view direction.} We compared the novel view synthesis results. The results without rotate the view direct would lead to wrong color expression.In test viewpoints, the spherical harmonic functions appear devoid of color, underscoring the importance of rotating view director with the rotation and aligning with the camera coordinate.}
  \label{fig:abrot}
\end{figure}
\subsection{Ablation Study}
We study the effect of various components of our method on the PeopleSnapshot dataset and the ZJU-MoCap dataset, including SMPL parameter refinement, the Rigid-based prior, and split with the scale. The average metrics over 4 sequences are reported in Tab.\ref{table:ablition study}. All proposed techniques are required to reach optimal performance.
% \noindent \textbf{Effective of Pose-guided deformation.}
% Our method extends the 3d Gaussian Splatting into the animatable avatar reconstruction, as shown in Table\ref{table:compare}, we modify the original 3D-GS\cite{kerbl20233d} with rotate the scene space with the global orient of SMPL parameter to suit the frozen camera, after such operate, gaussian still could not reconstruct the scene without the pose guided-deformation, which do not recover the 3d model, just recover a texture in the picture space, even 

\noindent \textbf{Effective of Rigid-based prior.}
To evaluate the impact of the Rigid-based prior, we conducted experiments by training models with part-specific Rigid-based priors. As shown in Table~\ref{table:ablition study}, our full model would gain the best performance among the recent approaches. The absence of prior challenges for the model in maintaining consistency across various viewpoints and movements in Fig.~\ref{fig:rigid-loss}. This is attributed to overfitting during training, the Gaussians would only fit part of the views in the canonical space, resulting in an inadequate representation of Gaussians during changes in pose and viewpoint to the observation space.

\noindent \textbf{Effective of rotate view direction}
As shown in the Fig.~\ref{fig:abrot}, it becomes apparent that without rotating the implementation direction, correct colors and results cannot be attained when rendering from alternative viewpoints. In test viewpoints, the spherical harmonic functions appear devoid of color, underscoring the importance of rotating view director with the rotation and aligning with the camera coordinate. This approach facilitates the accurate learning of colors and expressions by the Gaussian model.
% \begin{figure}
%         \centering
%         \includegraphics[width=1\linewidth]{pic_draft/sps.png}
%         \caption{\textbf{Effect of splitting with the scale.} The results without this refinement might generate artifacts in novel pose synthesis.}
%         \label{fig:split}
% \end{figure}
%

\noindent \textbf{Effective of Split-with-scale.}
% We compare it with the original 3D-GS split strategy with the threshold of gradient, constraining the gradient threshold still could not generate as good (dense\&smooth) results as our split-with-scale strategy in Figure.~\ref{fig:split_grad}. As monocular datasets lack sufficient variability in viewpoints, fewer and larger Gaussians are utilized in regions with minimal changes in motion and limited texture diversity. Consequently, the model tends to incorporate texture information into the geometric domain. More geometric details would be preserved in a more dense point cloud which is reflected in the metrics presented in Table~\ref{table:ablition study}. By employing additional splitting, the point cloud on the surface becomes enriched to present more geometry details. 
We compare our method with the original 3D-GS split strategy, which relies on a gradient threshold. Even when constraining the gradient threshold, the original strategy does not generate results as dense and smooth as our strategy, as shown in Figure~\ref{fig:split_grad}. In monocular datasets, which lack sufficient variability in viewpoints, fewer and larger Gaussians are used in regions with minimal motion changes and limited texture diversity. This causes the model to incorporate texture information into the geometric domain. A denser point cloud preserves more geometric details, as reflected in the metrics in Table~\ref{table:ablition study}. By employing additional splitting, our approach enriches the point cloud on the surface, capturing more geometric details. Despite our  strategy increasing the training parameters, our model still achieves comparable fast training and rendering speed.
% while improving the rendering quality by a large margin.

\noindent \textbf{Effective of Joint optimization of SMPL parameters.}
We utilize the SMPL model as the guide for rigid and non-rigid deformations, but inaccurate SMPL estimation could lead to inconsistencies of body parts in spatial positions under different viewpoints, resulting in blurred textures and floating artifacts, as shown in the metrics decrease in the Tab.\ref{table:ablition study} and worse visual in the Figure.~\ref{fig:refine}, such as the floating artifact on the sleeve and the blur texture on the collar.

\noindent \textbf{Effective of initial with SMPL vertex.} We initialize the canonical 3D Gaussians with vertex(N = 6890) of SMPL mesh in canonical pose. This enables us to generate suitable models in a relatively short time, allowing more time for subsequent texture mapping and pose optimization. We are able to produce models of decent quality in approximately 11 minutes and generate high-quality models within 20 to 30 minutes as shown in the Fig.\ref{fig:time}.
\section{Conclusion}
In this paper, we present AniGaussian, a novel method for reconstructing dynamic animatable avatar models from monocular videos using the 3D Gaussians Splatting representation. By incorporating pose guidance deformation with rigid and non-rigid deformation, we extend the 3D-GS representation to animatable avatar reconstruction. Then, we incorporate pose refinement to ensure clear textures. To mitigate between the observation space and the canonical space, we employ a rigid-based prior to regularizing the canonical space Gaussians and a split-with-scale strategy to enhance both the quality and robustness of the reconstruction. Our method is capable of synthesizing an animatable avatar that can be controlled by a novel motion sequences. Experiments on the PeopleSnapShot and ZJU-MoCap datasets, our method achieves superior quality metrics with the benchmark, demonstrating competitive performance.

\noindent \textbf{Future Work}
While our method is capable of producing high-fidelity animatable avatar and partially restoring clothing wrinkles and expressions, we have identified certain challenges. During training, Gaussians hard to assimilate texture colors into their internal representations, leading to difficulties in accurately learning surface details and textures with monocular input. Moreover, due to the inherent sensitivity of spherical harmonic functions to lighting, diffuse color and lighting would be baked into the spherical harmonic functions. In our future endeavors, we aim to decouple lighting and color expressions by leveraging different dimensions of spherical harmonic functions for separate learning. This approach will facilitate the generation of lighting decoupled and animatable avatar within a single learning stage.

\bibliographystyle{IEEEtran}
\bibliography{references}

\newpage

\begin{IEEEbiography}
[{\includegraphics[width=1in,height=1.25in,clip,keepaspectratio]{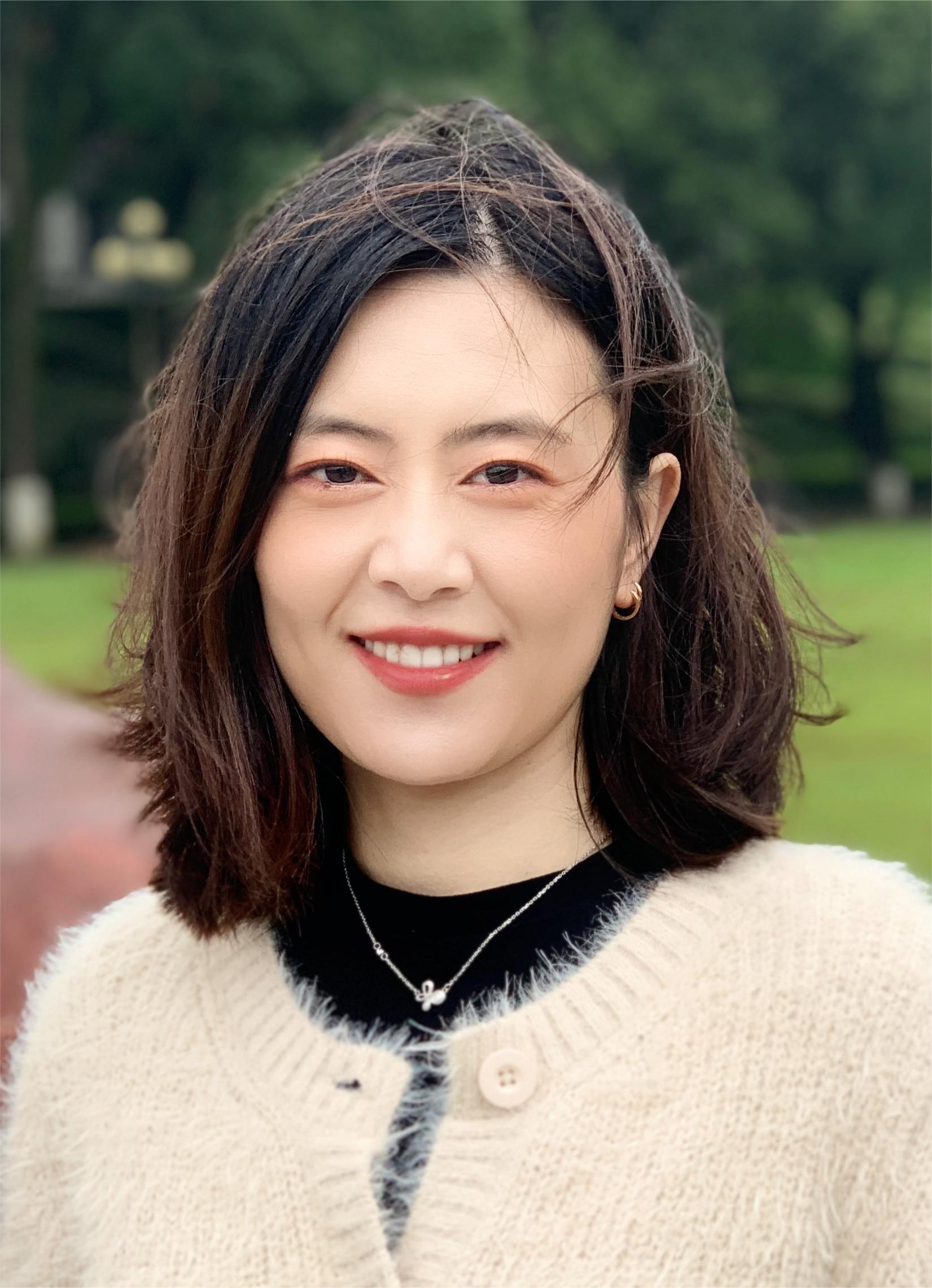}}]{Mengtian Li} currently hold a position as a Lecturer of Shanghai University, while simultaneously fulfilling the responsibilities of a Post-doc of Fudan University. She received Ph.D. degree from East China Normal University, Shanghai, China, in 2022. She serves as reviewers for CVPR, ICCV, ECCV, ICML, ICLR, NeurIPS , IEEE TIP and PR, etc. Her research lies in 3D vision and computer graphics, focuses at human avatar animating and 3D scene understanding, reconstruction, generation.
\end{IEEEbiography}

% \begin{IEEEbiography}
% [{\includegraphics[width=1in,height=1.25in,clip,keepaspectratio]{photo/zhai.jpg}}]{Chengshuo Zhai} received her Bachelor's degree in Digital Media Technology from Yanbian University in 2018 and is currently pursuing a Master's degree at the Shanghai Film Academy, Shanghai University. Her primary research focus is on human motion, specifically the generation of cross-modal human motion sequences.
% \end{IEEEbiography}
\vspace{-33pt}
\begin{IEEEbiography}
[{\includegraphics[width=1in,height=1.25in,clip,keepaspectratio]{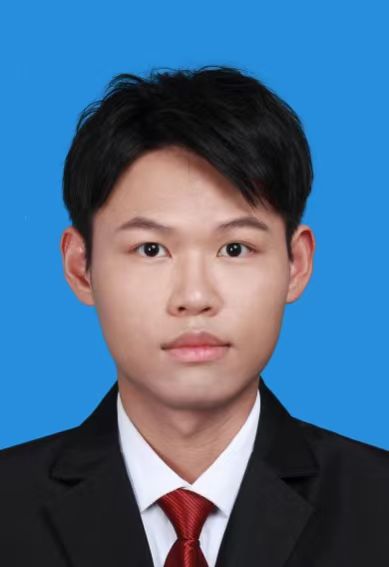}}]{Yaosheng Xiang} received his Bachelor's degree from Huaqiao University and is currently pursuing a Master's degree at the Shanghai Film Academy, part of Shanghai University. His research interests primarily focus on digital human reconstruction, with an emphasis on creating and editing animatable avatars from video footage.
\end{IEEEbiography}
\vspace{-33pt}
\begin{IEEEbiography}
[{\includegraphics[width=1in,height=1.25in,clip,keepaspectratio]{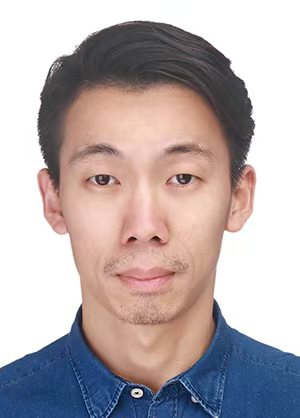}}]{Chen Kai} is currently a graduate supervisor at the Shanghai Film Academy of Shanghai University. He is the Director of the Shanghai Film Special Effects Engineering Technology Research Center and the Director of the Shanghai University film-producing workshop. He obtained a Master of Fine Arts (MFA) degree from the École Nationale Supérieure des Beaux-Arts de Le Mans in France, majoring in Contemporary Art. He participated in developing animation software Miarmy won the 70th Tech Emmy Awards presented by NATAS (National Academy of Television Arts \& Sciences) in 2018. His creative pursuits include experimental cinema, photography, digital interactive installations, and other forms of art.
\end{IEEEbiography}
\vspace{-33pt}
\begin{IEEEbiography}
[{\includegraphics[width=1in,height=1.25in,clip,keepaspectratio]{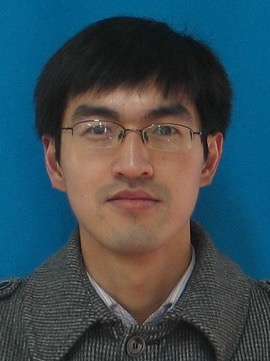}}]{Zhifeng Xie} received the Ph.D. degree in computer application technology from Shanghai Jiao Tong University, Shanghai, China. He was a Research Assistant with the City University of Hong Kong, Hong Kong. He is currently an Associate Professor with the Department of Film and Television Engineering, Shanghai University, Shanghai. He has published several works on CVPR, ECCV, IJCAI, IEEE Transactions on Image Processing, IEEE Transactions on Neural Networks and Learning Systems, and IEEE Transactions on Circuits and Systems for Video Technology. His current research interests include image/video processing and computer vision.
\end{IEEEbiography}
\vspace{-33pt}
\begin{IEEEbiography}[{\includegraphics[width=1in,height=1.25in,clip,keepaspectratio]{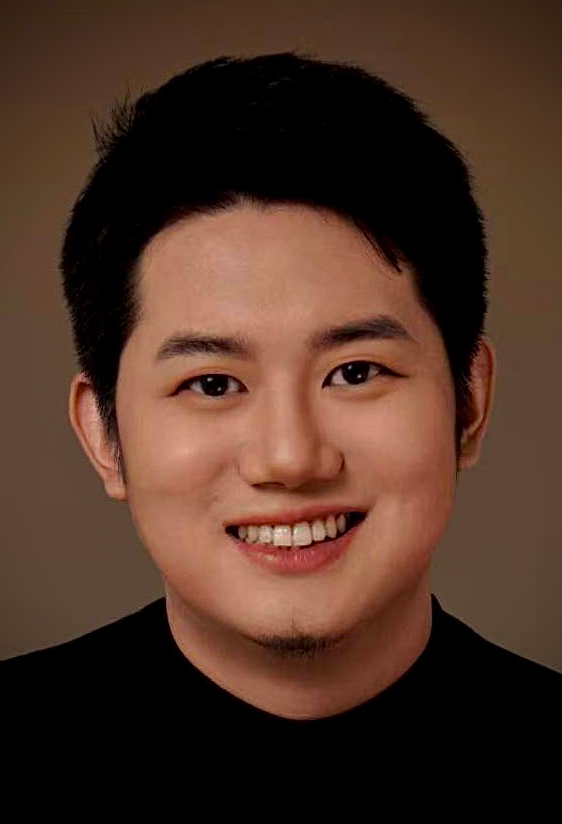}}]{Keyu Chen} is a senior AI researcher affiliated with Tavus Inc.. He obtained the master and bachelor degree from University of Science and Technology of China in 2021 and 2018. His research interests are mainly focused on digital human modeling, animation, and affective analysis.
\end{IEEEbiography}

\vspace{-33pt}
\begin{IEEEbiography}
[{\includegraphics[width=1in,height=1.25in,clip,keepaspectratio]{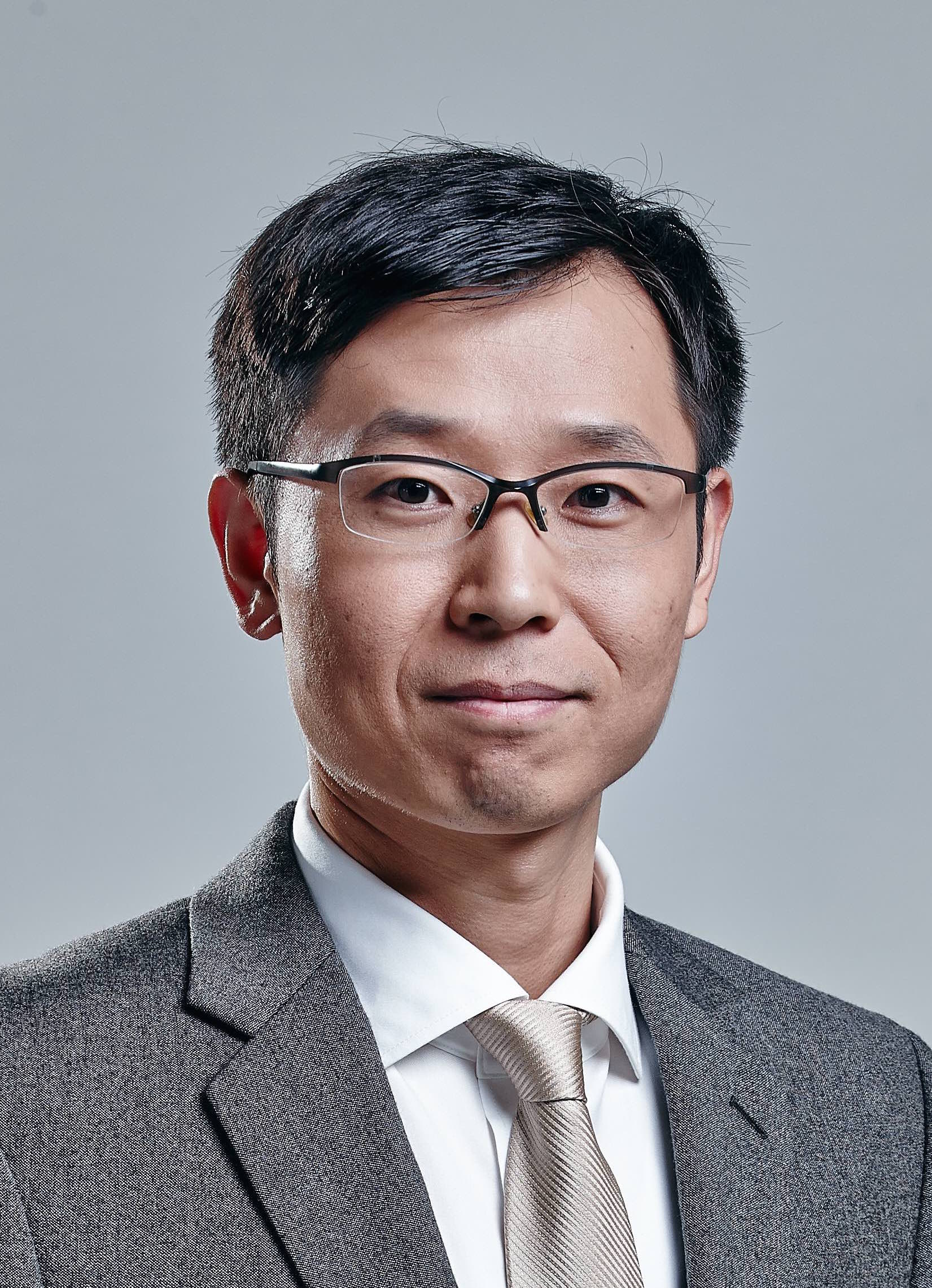}}]{Yu-Gang Jiang} received the Ph.D. degree in Computer Science from City University of Hong Kong in 2009 and worked as a Postdoctoral Research Scientist at Columbia University, New York, from 2009 to 2011. He is currently Vice President and Chang Jiang Scholar Distinguished Professor of Computer Science at Fudan University, Shanghai, China. His research lies in the areas of multimedia, computer vision, and trustworthy AGI. He is a fellow of the IEEE and the IAPR.
\end{IEEEbiography}

\end{document}